\documentstyle[12pt]{article}
\begin{document}


\def\b #1{\mbox{\footnotesize\boldmath$#1$\normalsize}}

\def\bb#1{\mbox{\footnotesize\boldmath$\overline{#1}$\normalsize}}

\def\SG{\mbox{\Large$\sigma$\normalsize}}

\def\DG{\mbox{\large$\delta$\normalsize}}

\title{On the extended loop calculus\footnote{To appear in
``Proceedings of the LASSF II, Theoretical Physics'$\,95$", Ed. by A.
Restuccia et al., U.S.B. (Caracas).}}

\author{\small Jorge Griego
\\ \small Instituto de F\'{\i}sica, Facultad de Ciencias,
\\ \small Trist\'an Narvaja 1674, 11200 Montevideo, Uruguay.
\\ \small E-mail: griego@fisica.edu.uy}

\date{December 4, 1995}
\maketitle
\abstract{Some features of extended loops are considered. In particular,
the behaviour under diffeomorphism transformations of the wavefunctions
with support on the extended loop space  are studied. The basis of a
method to obtain analytical expressions of diffeomorphism invariants
via extended loops are settled. Applications to knot theory and quantum
gravity are considered.}

\section{Introduction}
Extended loops arise as generalizations of ordinary loops
\cite{DiGaGr1}. The set of all extended loops can be viewed as a
manifold. This manifold can be endowed with a (local) infinite
dimensional Lie group structure that contains the usual group of loops
as a discrete subgroup. In the same way that in the case of ordinary
loops, any Lie-algebra valued connection theory can be transcribed to
the language of extended loops. As the group of extended loops has a
more rich mathematical structure than the conventional loop space,
several benefits at the calculation and regularization levels are
exhibited by the extended loop representation. These advantages has been
proved to be relevant in the study of the space of states of quantum
gravity \cite{DiGaGr2,DiGaGrPu2,Gr}.

The purpose of this talk is to give a primary approach to the extended
loop calculus, specially attending its applications to knot theory and
quantum gravity. We analyze with some detail the mechanism of operation
of the generator of diffeomorphism transformations on the extended loop
wavefunctions and settle the basis of a systematic method to
obtain analytic expressions of knot invariants using extended
loops. The significance of the method (that provide analytical
expression of knot invariants in a generic way) for the case of quantum
gravity are discussed.

The next section contains several preliminary remarks concerning extended
loops. In Sect. 3 the properties of the wavefunctions defined in the
extended loop space are considered. In Sect. 4 the mechanism of
operation of the generator of diffeomorphism transformations over the
extended loop wavefunctions is analyzed. In Sect. 5 a prescription to
build up ``families" of extended knot invariants is developed and
applied in some simple cases. The conclusions are in Sect. 6.

\section{Preliminaries}
The use of extended loops requires a new methodology. In some sense
this new way to look loops is present at the level of ordinary loops. In
fact, it was the intention to solve some problems of the conventional
loop representation that had lead to the discovery of the
extended loop group. In order to fix a point of departure towards the
extended loop world we sketch briefly the history of the primary steps.

Holonomies are the basic quantities that underlies any loop
representation of a Lie-algebra valued connection theory. They are given
by the ordered exponential of a line integral of the connection along a
close curve $\gamma$:

\begin{eqnarray}
H_A (\gamma)&:=& \em{P} \exp{\oint_\gamma d y^a A_a(y)}=
1+\oint_\gamma d y^a A_a(y)+\cdots+
\nonumber\\
&&\oint_\gamma dy_r^{a_r}\int_o^{y_r} dy_{r-1}^{a_{r-1}}\ldots
\int_o^{y_2} dy_1^{a_1} A_{a_1}(y_1)\ldots A_{a_r}(y_r)+ \cdots
\label{holo}
\end{eqnarray}
We write

\begin{eqnarray}
\oint_\gamma d y^a A_a(y)&=&\oint_\gamma d y^a \int d^3 x \delta(x-y)
A_a(x)=\int d^3 x A_a(x) \oint_\gamma d y^a \delta(x-y)
\nonumber\\
&:=&\int d^3 x A_a(x) X^{a}(x,\gamma)
\equiv A_{ax}\, X^{ax}(\gamma)
\label{rank1}
\end{eqnarray}
$X^{ax}(\gamma)$ is the {\it multitangent} field of {\it rank} one
associated with the loop $\gamma$. This field is a loop dependent
distribution that admits a direct geometrical interpretation: it fixes
the tangent at the point $x$. Notice that in (\ref{rank1}) the indices
were grouped into pairs and a generalized Einstein convention was used.
We write the paired indices with Greek letters, $\mu:=ax$. This notation
is in agreement with the behaviour of the multitangents under general
coordinate transformations\footnote{The multitangents behave as
generalized multitensors with respect to the Greek indices.}. Repeating
the procedure with the other terms in (\ref{holo}) we get

\begin{eqnarray}
H_A (\gamma)&=&1+ A_{\mu_1}\,X^{\mu_1}(\gamma)+\cdots+
A_{\mu_1\ldots\mu_r}\,X^{\mu_1\ldots\mu_r}(\gamma)+\cdots
\nonumber\\
&:=&A_{\b\mu}\,X^{\b\mu}(\gamma)
\end{eqnarray}
$X^{\mu_1\ldots\mu_r}(\gamma)$ is the multitangent field of rank
$r(\b\mu)=r$ and it is given by the ordered integration along the loop
of the product of $r$ delta functions. In general a bold Greek index
represents an ordered set of paired indices $\b\mu:=\mu_1\ldots\mu_r$
and repeated bold Greek indices indicates a summation from rank
zero\footnote{We use the convention $X^{\mu_1\ldots\mu_0}\equiv 1$.} to
infinity. By this procedure one is capable to write the holonomy in
a very compact and useful form. In particular one sees that all the
information about $\gamma$ needed to construct the holonomy is contained
in the multitangents of all ranks. As far as the loop representation is
concered ordinary loops can then be viewed as entirely equivalent to the
infinite string of multitangent fields:

\begin{equation}
\gamma \leftrightarrow {\bf X}(\gamma):=(1,X^{\mu_1}(\gamma)+\cdots+
X^{\mu_1\ldots\mu_r}(\gamma)+\cdots)
\end{equation}
Extended loops arise as generalizations of the multitangents in order
to include more general fields. In general an extended loop will be
given by the string

\begin{equation}
{\bf X}:=(X,X^{\mu_1}+\cdots+
X^{\mu_1\ldots\mu_r}+\cdots)\equiv
(X^{\b\mu}\,;\;\,r(\b\mu)=0,\ldots,\infty)
\end{equation}
where $X$ is a real number and the $X^{\mu_1\ldots\mu_r}$ satisfy
several properties inherited from the multitangents. These properties
are:

\begin{description}
\item{\it i.} Coordinate transformations properties:
$X^{\mu_1\ldots\mu_r}$ behaves as a vector density with respect to each
``contravariant" index.

\item{\it ii.} The differential constraint: the divergence of a
multivector of rank $r$ generates a multivector of rank $r-1$ (this
property makes holonomies formally covariant under gauge transformations).

\item{\it iii.} The algebraic constraint: the linear combination
$X^{\underline{\b\alpha}\b\beta}$ of multivectors of rank
$r(\b\alpha)+r(\b\beta)$ where all the indices are permuted {\it
preserving} the relative order of the set $\b\alpha$ and $\b\beta$ split
into the product ${\textstyle{1\over X}}X^{\b\alpha}\,X^{\b\beta}$ (this
property is related to the existence of an ``order" for the
contravariant indices of the multitangents -originated by the ordered
integration of the distributions along the loop-).

\end{description}
The set $\{{\bf X}\}$ of all extended loops constitutes a vector space
that can be endowed with the structure of a (local) infinite dimensional
Lie group by means of the following composition law:

\begin{equation}
[{\bf X}_1\times {\bf X}_2]^{\b \mu} :=
\DG^{\b \mu}_{\b \pi \b \theta} \, X_1^{\b \pi}\,
X_2^{\b \theta}
\end{equation}
where\footnote{$\delta^{\mu_i}_{\nu_i}:=\delta^{a_i}_{b_i}\delta(x_i-y_i)$,
with $\mu_i=a_i x_i$ and $\nu_i=b_i y_i$.}

\begin{equation}
\DG^{\b \mu}_{\b \nu} := \left\{
\begin{array}{ll}
\delta^{\mu_1}_{\nu_1} \cdots \delta^{\mu_r}_{\nu_r} \;, &
\mbox{ if $r({\b \mu}) = r({\b \nu}) = r \geq 1$} \rule{0mm}{6mm}
\\ 1 \;, &\mbox{ if $r({\b \mu}) = r({\b \nu}) = 0$}
\rule{0mm}{6mm} \\ \rule{0mm}{6mm} 0\;, &\mbox{ in other case}
\end{array}
\right. \label{deltag}
\end{equation}
The $\DG$-matrix is a very useful device to compact  sums of multivector
fields. The above definition of the extended group product is the
compact version of the following sum:

\begin{equation}
[{\bf X}_1\times {\bf X}_2]^{\mu_1\ldots\mu_r} =
\sum_{k=0}^{r}\, X_1^{\mu_1\ldots\mu_k}\,
X_2^{\mu_{k+1}\ldots\mu_r}
\end{equation}
where now $X^{\mu_{j+1}\ldots\mu_j}=X$ gives the component of rank zero
of the extended loop. Notice that the $\DG$-matrix allows to write an
extended loop in the following way\footnote{This notation is due to C. Di
Bartolo.}

\begin{equation}
{\bf X}:={\b \DG}_{\b \nu}\,X^{\b \nu}
\end{equation}
The set of ``covariant" multivectors ${\b \DG}_{\b \nu}$ can be viewed
as a basis of the vector space $\{{\bf X}\}$. In order to connect
extended with ordinary loops we use the following properties of the
multitangents:

\begin{eqnarray}
{\bf X}(\gamma_1)\times {\bf X}(\gamma_2)&=&{\bf X}(\gamma_1\gamma_2)
\\
X^{\b\mu}(\overline{\gamma})&=&X^{\bb\mu}(\gamma)
\end{eqnarray}
where $\gamma_1\gamma_2$ is the group product in the nonparametric loop
space, $\overline{\gamma}$ is the rerouted loop and
$\bb\mu:=(-1)^{r(\b\mu)}\b\mu^{-1}$ with
$\b\mu^{-1}:=\mu_r\ldots\mu_1$.

Extended loops and ordinary loops are then closely related by the
multitangent fields. This relationship is also put into manifest at the
level of the representations: through the use of the multitangents, the
holonomy admits a direct generalization to the extended loop space. An
extended holonomy will be defined by the series

\begin{equation}
H_A ({\bf X}):=A_{\b\mu}\,X^{\b\mu}
\end{equation}
and they allow to formally represent any Lie-algebra valued connection
theory in the extended loop space in close resemblance with the case of
ordinary loops\footnote{Extended holonomies are affected by convergence
problems that question the gauge invariance of the
representation\cite{Troy}. There exist several alternatives to solve
this problem, as the one presented by C. Di Bartolo \cite{Di} in this
volume. See also \cite{DiGaGrPu2}.}. In what follows we shall study the
diffeomorphism transformation properties of the wavefunctions in the
extended loop representation.

\section{Extended knot invariants}
Linear extended loop wavefunctions are written in the following general
form

\begin{equation}
\psi({\bf X})=\psi_{\b\mu}\,X^{\b\mu}
\label{lwf}
\end{equation}
where the propagators $\psi_{\b\mu}$ characterizes completely the state
$\psi$. Any extended loop wavefunction generates a loop wavefunction
when the multivectors are particularized to the multitangents:

\begin{equation}
\psi({\bf X})\rightarrow\psi(\gamma)=\psi_{\b\mu}\,X^{\b\mu}(\gamma)
\end{equation}
The converse in not true in general. As it was mentioned, the components
of an extended loop behave as multivector densities under general
coordinate transformations. Using this fact\footnote{Another possibility
is by translating the generator of diffeomorphisms from the
space of connections to the space of extended loops, see
\cite{DiGaGr2}.} it is possible to derive the following transformation
law for the extended loop wavefunction under infinitesimal coordinate
transformations $x'^a=x^a + \eta^a (x)$:

\begin{equation}
\psi({\bf X}')=\psi({\bf X})+\eta^{ax}\,{\cal C}_{ax}\,\psi({\bf X})
\label{difwf}
\end{equation}
where the generator of diffeomorphisms is given by a linear expression
in the functional derivatives of the multivector fields\footnote{Notice
that the action of the diffeomorphism operator reduces to a shift of the
argument of linear wavefunctions.}:

\begin{equation}
{\cal C}_{ax}\,\psi({\bf X})
:=[{\cal F}_{ab}(x)\times{\bf X}^{(bx)}]^{\b
\mu}\frac{\delta}{\delta\,X^{\b \mu}}\,\psi({\bf X})
\equiv\psi_{\b\alpha\b\beta}\,{\cal F}_{ab}^{\b
\alpha}(x)\,X^{(bx\,\b \beta)_c}
\label{exdif}
\end{equation}
In the above expression,

\begin{equation}
{\cal F}_{ab}^{\alpha_1\ldots\alpha_r}(x):=\epsilon_{abc}\,[-
\DG^{\alpha_1\ldots\alpha_r}_{\nu_1}\,g^{cx\,\nu_1}+
\DG^{\alpha_1\ldots\alpha_r}_{\nu_1\nu_2}\,\epsilon^{cx\,\nu_1\nu_2}]\;,
\label{fab}
\end{equation}

\begin{equation}
g^{cx\,\nu_1}:=\epsilon^{cb_1k}\,\partial_k\,\delta(x-y_1)\;,
\end{equation}
and

\begin{equation}
\epsilon^{cx\,\nu_1\nu_2}:=\epsilon^{cb_1b_2}
\delta(x-y_1)\delta(x-y_2)
\end{equation}
In the last expressions a mixed notation of Greek and paired indices
were used. In the case that extended loops are particularized to
ordinary loops we can write from (\ref{exdif})

\begin{eqnarray}
{\cal C}_{ax}\,\psi[{\bf X}(\gamma)]&=&\psi_{\b\alpha\pi\b\theta}
\,{\cal F}_{ab}^{\b\alpha}(x)\,X^{\b\theta\,bx\,\b\pi}(\gamma)
\nonumber\\
&=&\oint_{\gamma}\,dy^b \delta(x-y)\psi_{\b\alpha\b\pi\b\theta}
\,{\cal F}_{ab}^{\b\alpha}(y)\,
X^{\b\pi}(\gamma_y^o)\,X^{\b\theta}(\gamma_o^y)
\nonumber\\
&=&\oint_{\gamma}\,dy^b \delta(x-y)\psi_{\b\mu}
\,[{\cal F}_{ab}(x)\times{\bf X}(\gamma_y^o)\times{\bf
X}(\gamma_o^y)]^{\b\mu}
\label{paso}
\end{eqnarray}
where $\gamma_o^y$ is the portion of the loop from the origin $o$ to the
point $y$. Introducing the identity $1={\bf
X}(\overline{\gamma}_o^y)\times {\bf X}(\gamma_o^y)$ and using the
cyclicity of the set $\b \mu$\footnote{The fact that $\psi_{\b\mu}$ is
invariant under a cyclic permutation of the indices assures that the
wavefunctions do not depend of the origin $o$ of the loops.} we get from
(\ref{paso})

\begin{eqnarray}
{\cal C}_{ax}\psi(\gamma)&=&
\oint_{\gamma}\,dy^b \delta(x-y)\psi_{\b\mu}
[{\bf X}(\gamma_o^y)\times{\cal F}_{ab}(y)\times
{\bf X}(\overline{\gamma}_o^y)\times{\bf X}(\gamma_o^y)\times
{\bf X}(\gamma_y^o)]^{\b\mu}
\nonumber\\
&\equiv&
\oint_{\gamma}\,dy^b\, \delta(x-y)\,\Delta_{ab}(\gamma^y)
\,\psi(\gamma)
\label{diflw}
\end{eqnarray}
where

\begin{equation}
\Delta_{ab}(\gamma^y)\,{\bf X}(\gamma):=
[{\bf X}(\gamma_o^y)\times{\cal F}_{ab}(y)\times
{\bf X}(\overline{\gamma}_o^y)]\times{\bf X}(\gamma)
\label{loopder}
\end{equation}
is the loop derivative \cite{GaTr} defined in the nonparametric loop space.
This result shows that the expression (\ref{exdif}) gives the correct
transformation law under diffeomorphisms for ordinary loop wavefunctions
once the multivectors are specialized to the multitangents. This
means that any solution of the equation

\begin{equation}
{\cal C}_{ax}\,\psi({\bf X}) = 0
\label{eki}
\end{equation}
can be viewed as an ``extended knot invariant" (in the sense that the
restriction of the domain of definition of the solution to ordinary
loops will always generate a knot invariant). According to (\ref{lwf})
and (\ref{fab}) we have

\begin{equation}
{\cal C}_{ax}\,\psi({\bf X}) =
\sum_{r=0}^{\infty} \epsilon_{abc}\psi_{\mu_1 \ldots \mu_r}[-
g^{cx\,\mu_1}\,X^{(bx\,\mu_2 \ldots \mu_r)_c}+
\epsilon^{cx\,\mu_1 \mu_2}\,X^{(bx\,\mu_3 \ldots \mu_r)_c}]=0
\label{difwf2}
\end{equation}
In the following section we analyze how (\ref{difwf2}) works.

\section{The action of ${\cal C}_{ax}$}
The possibility to systematize the search of solutions of (\ref{difwf2})
is based on the following observation: for all the known
analytical expressions of knot invariants, the propagators
$\psi_{\mu_1\ldots\mu_r}$ are completely expressed in terms of the two
($g_{\cdot\cdot}$) and three ($h_{\cdot\cdot\cdot}$) point propagators
of the Chern-Simons theory, given by

\begin{equation}
g_{\mu_1\mu_2}\equiv\epsilon_{a_1 a_2 k}\phi^{\,kx_1}_{\,x_2}:=
-\epsilon_{a_1 a_2 k}\frac{\partial_k}{\nabla^2}\delta(x_1-x_2)
\label{g}
\end{equation}
and

\begin{equation}
h_{\mu_1\mu_2\mu_3}:=\epsilon^{\alpha_1 \alpha_2 \alpha_3}
g_{\mu_1\alpha_1}g_{\mu_2\alpha_2}g_{\mu_3\alpha_3}
\label{h}
\end{equation}
Moreover, the following properties are also used:

\begin{enumerate}
\item The propagators $\psi_{\mu_1\ldots\mu_r}$ {\it vanishes} for a certain
(maximum) rank $N$; that is to say, $\psi_{\mu_1\ldots\mu_r}=0$ for all
$r>N$.

\item For $r=N$, $\psi_{\mu_1\ldots\mu_N}$ is given exclusively by
products of two point Chern-Simons propagators.

\item The minimum rank $n$ does not contain any two point Chern-Simons
propagator.

\end{enumerate}
The successive ranks of the wavefunction are linked by the action of the
operators. Let us consider the case of $N=6$ and $n=4$. The general
form of the wavefunctions of this type is:

\begin{equation}
\psi({\bf X})=g_{\cdot\cdot}g_{\cdot\cdot}g_{\cdot\cdot}
X^{\cdot\cdot\cdot\cdot\cdot\cdot}
+h_{\cdot\cdot\cdot}g_{\cdot\cdot}
X^{\cdot\cdot\cdot\cdot\cdot}
+h_{\cdot\cdot\star}g^{\star\star}h_{\star\cdot\cdot}
X^{\cdot\cdot\cdot\cdot}
\end{equation}
and one obtains the following general picture for the result of the
application of ${\cal C}_{ax}$ onto this state:
\vspace{0.5cm}

\footnotesize
\begin{eqnarray*}
{\cal C}_{ax}\,g_{\mbox{\boldmath$\cdot\cdot$}}
g_{\mbox{\boldmath$\cdot\cdot$}}g_{\mbox{\boldmath$\cdot\cdot$}}
X^{\mbox{\boldmath$\cdot$}\mbox{\boldmath$\cdot$}
\mbox{\boldmath$\cdot$}\mbox{\boldmath$\cdot$}\mbox{\boldmath$\cdot$}
\mbox{\boldmath$\cdot$}}\!\!
&:&\!\!-\epsilon_{abc}
g_{\mbox{\boldmath$\cdot\cdot$}}
g_{\mbox{\boldmath$\cdot\cdot$}}
X^{(bx\,\mbox{\boldmath$\cdot\cdot$}\,cx\,\mbox{\boldmath$\cdot\cdot$})_c}
+G_{ax\,bx\,\mbox{\boldmath$\cdot$}\mbox{\boldmath$\cdot$}
\mbox{\boldmath$\cdot$}\mbox{\boldmath$\cdot$}}
X^{(bx\,\mbox{\boldmath$\cdot$}\mbox{\boldmath$\cdot$}
\mbox{\boldmath$\cdot$}\mbox{\boldmath$\cdot$})_c}
\\ \\
{\cal C}_{ax}\,h_{\mbox{\boldmath$\cdot$}\mbox{\boldmath$\cdot$}
\mbox{\boldmath$\cdot$}}g_{\mbox{\boldmath$\cdot\cdot$}}
X^{\mbox{\boldmath$\cdot$}\mbox{\boldmath$\cdot$}
\mbox{\boldmath$\cdot$}\mbox{\boldmath$\cdot$}
\mbox{\boldmath$\cdot$}}\!\!&:&\!\!
-\epsilon_{abc}
h_{\mbox{\boldmath$\cdot$}\mbox{\boldmath$\cdot$}\mbox{\boldmath$\cdot$}}
X^{(bx\,\mbox{\boldmath$\cdot$}\,cx\,\mbox{\boldmath$\cdot\cdot$})_c}
-G_{ax\,bx\,\mbox{\boldmath$\cdot$}\mbox{\boldmath$\cdot$}
\mbox{\boldmath$\cdot$}\mbox{\boldmath$\cdot$}}
X^{(bx\,\mbox{\boldmath$\cdot$}\mbox{\boldmath$\cdot$}
\mbox{\boldmath$\cdot$}\mbox{\boldmath$\cdot$})_c}
+H_{ax\,bx\,\mbox{\boldmath$\cdot$}\mbox{\boldmath$\cdot$}
\mbox{\boldmath$\cdot$}}
X^{(bx\,\mbox{\boldmath$\cdot$}\mbox{\boldmath$\cdot$}
\mbox{\boldmath$\cdot$})_c}
\\  \\
{\cal C}_{ax}\,h_{\mbox{\boldmath$\cdot$}\mbox{\boldmath$\cdot$}\star}
g^{\star\star}
h_{\star\mbox{\boldmath$\cdot$}\mbox{\boldmath$\cdot$}}
X^{\mbox{\boldmath$\cdot$}\mbox{\boldmath$\cdot$}
\mbox{\boldmath$\cdot$}\mbox{\boldmath$\cdot$}}\!\!&:&\!\!
-H_{ax\,bx\,\mbox{\boldmath$\cdot$}\mbox{\boldmath$\cdot$}
\mbox{\boldmath$\cdot$}}
X^{(bx\,\mbox{\boldmath$\cdot$}\mbox{\boldmath$\cdot$}
\mbox{\boldmath$\cdot$})_c}
+I_{ax\,bx\,\mbox{\boldmath$\cdot$}\mbox{\boldmath$\cdot$}}
X^{(bx\,\mbox{\boldmath$\cdot$}\mbox{\boldmath$\cdot$})_c}
\end{eqnarray*}
\normalsize
\vspace{0.5cm}

\noindent
$G$, $H$ and $I$ are certain expressions containing $g$'s and/or $h$'s
with two spatial indices fixed at the point $x$. Notice that $G\cdot X$
represents a contribution of rank five, $H\cdot X$ one of rank four and
$I\cdot X$ one of rank three. As a general rule, the action of ${\cal
C}_{ax}$ on a rank $r$ of the wavefunction generates two types of
contributions, one of rank $r$ and other of rank $r-1$. For $r>n$ the
contribution of rank $r-1$ is always canceled by terms that appear when
the operator acts on the rank $r-1$. A chain of cancellations linking
intimately the successive ranks of the wavefunction then occurs induced
by the diffeomorphism generator \footnote{Exactly the same chain of
cancellations takes place also in the case of the Hamiltonian constraint
of quantum gravity.}. Two more punctuations can be quoted from the above
result:

\begin{description}
\item{P1:} For $n<r\leq N$ a contribution of the form
$\epsilon_{abc}\psi_{\ldots\ldots}X^{(bx\,\ldots\,cx\,\ldots)_c}$
appears. This type of contributions do not enter in the chain of
cancellations and they have to vanish by means of symmetry
considerations\footnote{For extended knot invariants these terms would
involve in general a symmetric expression in the contravariant indices
$bx$ and $cx$.}.

\item{P2:} The term of lower rank is responsible of the closure of the
chain of cancellations in a consistent way; that is to say, the
contribution of rank $n-1$ {\it has to vanish
identically}\footnote{Notice that we do not dispose in this case of a
symmetry argument like in P1.}.

\end{description}
The systematic operation displayed by the diffeomorphism operator is
essentially a consequence of the fact that $g_{\cdot\cdot}$ and
$h_{\cdot\cdot\cdot}$ were taken as the building blocks of the
propagators $\psi_{\b\mu}$. It is worth to emphasize that no other
propagators than the Chern-Simons are known at present to participate in
the analytical expression of the knot invariants. In what follows we are
going to see that this procedure can be used to build up analytical
expressions of extended knot invariants in terms of $g$ and $h$ in a
generic way.

\section{Extended knot families}
A family of extended knot invariants is a set $\{\psi^{[N,n]}_i\}$ of
wavefunctions with the same maximum and minimum ranks that satisfy the
following properties:

\begin{description}
\item{F1:} ${\cal C}_{ax}\,\psi^{[N,n]}_i=0$ for all $i$.

\item{F2:} $\psi^{[N,n]}_i({\bf X})=\psi^{i}{_{\b\mu}}\,X^{\b\mu}$
with $\psi^{i}{_{\b\mu}}$ a cyclic propagator.

\item{F3:} $\psi^{i}{_{\mu_1\ldots\mu_n}}$ is the {\it same} for all
members of the family.

\end{description}
The following steps allow to construct extended knot families in a
systematic way:

\begin{description}
\item{$\underline{\mbox{Step 1:}}$} For the minimum rank $n$
construct all the cyclic combinations involving only three point
Chern-Simons propagators $h$ (and its contractions if necessary).

\item{$\underline{\mbox{Step 2:}}$} Identify those combinations that
satisfy the consistence condition P2. Each one of these combinations
$\psi^{i}_{\mu_1\ldots\mu_n}$ could be the origin of a family of knot
invariants.

\item{$\underline{\mbox{Step 3:}}$} Determine the cyclic combinations of
rank $n+1$ (where $h$'s are substituted by $g$'s) that makes

\begin{displaymath}
{\cal C}_{ax}\,\{\psi^{i}_{\mu_1\ldots\mu_n}X^{\mu_1\ldots\mu_n}+
\psi^{i}_{\mu_1\ldots\mu_{n+1}}X^{\mu_1\ldots\mu_{n+1}}\}
=\mbox{terms of rank $n+1$}
\end{displaymath}
Verify that the result does not include a remnant contribution of the
type described in P1.

\item{$\underline{\mbox{Step 4:}}$} Repeat the procedure for the
successive increasing ranks until all the $h$'s were replaced by $g$'s.

\end{description}
We see that the procedure of construction goes from the minimum to the
maximum rank. The most simple family has only one member, the Gauss
invariant:

\begin{equation}
\{\psi^{[2,2]}_1\equiv\varphi_G\}\;\;\mbox{\footnotesize{with}}\;\,
\varphi_G:=g_{\mu_1\mu_2}\,X^{\mu_1\mu_2}
\end{equation}
The next family has $N=4$ and $n=3$. Their members have the following
general from

\begin{equation}
\psi^{[4,3]}= g_{\cdot\cdot}g_{\cdot\cdot}X^{\cdot\cdot\cdot\cdot}+
h_{\cdot\cdot\cdot}X^{\cdot\cdot\cdot}
\end{equation}
We shall see how the method works in this simple case:

\begin{description}
\item{$\underline{\mbox{Step 1:}}$} $h_{\mu_1\mu_2\mu_3}$ is the only
possibility.

\item{$\underline{\mbox{Step 2:}}$}
\vspace{-0.4cm}

\begin{eqnarray}
\lefteqn{ {\cal C}_{ax}h_{\mu_1\mu_2\mu_3}X^{\mu_1\mu_2\mu_3}= }
\nonumber\\
&&\{-g_{\mu_1\,[ax\,}g_{bx]\,\mu_2}+\epsilon_{abc}(\phi^{\,cx}_{\,x_1}
-\phi^{\,cx}_{\,x_2})g_{\mu_1\mu_2}\}\,X^{(bx\,\mu_1\mu_2)_c}
\nonumber\\
&&+2\{h_{ax\,bx\,\mu_1}-\epsilon_{abc}\phi^{\,cx}_{\,z}\phi^{\,dz}_{\,x}
g_{\mu_1\,dz}\}\,X^{(bx\,\mu_1)_c}
\label{difh}
\end{eqnarray}
The consistency condition reads in this case:

\begin{equation}
I_{ax\,bx\,\mu_1}:=h_{ax\,bx\,\mu_1}-\epsilon_{abc}\phi^{\,cx}_{\,z}
\phi^{\,dz}_{\,x}g_{\mu_1\,dz}\equiv 0
\end{equation}
that is verified directly by developing $h_{ax\,bx\,\mu_1}$ according to
(\ref{h}).

\item{$\underline{\mbox{Step 3:}}$} The next (and last) rank includes
the following (independent) cyclic combinations of two $g$'s:

\begin{eqnarray}
&&C^1_{\mu_1\ldots\mu_4}=g_{\mu_1\mu_2}g_{\mu_3\mu_4}
+g_{\mu_1\mu_4}g_{\mu_2\mu_3}\\
&&C^2_{\mu_1\ldots\mu_4}=g_{\mu_1\mu_3}g_{\mu_2\mu_4}
\end{eqnarray}
The following results are obtained:
\vspace{-0.4cm}

\begin{eqnarray}
&&\hspace{-1cm}
{\cal C}_{ax}C^1_{\ldots}X^{\ldots}=
\{-g_{\mu_1\,[ax\,}g_{bx]\,\mu_2}+\epsilon_{abc}(\phi^{\,cx}_{\,x_1}
-\phi^{\,cx}_{\,x_2})g_{\mu_1\mu_2}\}X^{(bx\,\mu_1\mu_2)_c}
\\
&&\hspace{-1cm}
{\cal C}_{ax}C^2_{\ldots}X^{\ldots}=
\{g_{\mu_1\,[ax\,}g_{bx]\,\mu_2}-\epsilon_{abc}(\phi^{\,cx}_{\,x_1}
-\phi^{\,cx}_{\,x_2})g_{\mu_1\mu_2}\}X^{(bx\,\mu_1\mu_2)_c}
\end{eqnarray}
Notice that there are not terms of the form
$\epsilon_{abc}\psi_{\ldots\ldots}X^{(bx\,\ldots\,cx\,\ldots)_c}$. This
is due to the cyclicity of the $C$'s. Comparing with (\ref{difh}) we see
that there exist two possibilities to cancel the contribution of rank
three:

\begin{eqnarray}
&&\hspace{-0.6cm}\psi^{[4,3]}_1=(g_{\mu_1\mu_2}g_{\mu_3\mu_4}
+g_{\mu_1\mu_4}g_{\mu_2\mu_3})X^{\mu_1\mu_2\mu_3\mu_4}
-h_{\mu_1\mu_2\mu_3}X^{\mu_1\mu_2\mu_3}
\\
&&\hspace{-0.6cm}
\psi^{[4,3]}_2=g_{\mu_1\mu_3}g_{\mu_2\mu_4}X^{\mu_1\mu_2\mu_3\mu_4}
+h_{\mu_1\mu_2\mu_3}X^{\mu_1\mu_2\mu_3}
\end{eqnarray}
These are the members of the family $[4,3]$.
\end{description}
Notice that

\begin{eqnarray}
\psi^{[4,3]}_1+\psi^{[4,3]}_2&=&(g_{\mu_1\mu_2}g_{\mu_3\mu_4}
+g_{\mu_1\mu_3}g_{\mu_2\mu_4}
+g_{\mu_1\mu_4}g_{\mu_2\mu_3})X^{\mu_1\mu_2\mu_3\mu_4}
\nonumber\\
&=& \textstyle{1\over2}(\psi^{[2,2]}_1)^2
\end{eqnarray}
This linear expression corresponds to the square of the Gauss invariant
in the general extended space\footnote{In order to fulfill the linearity
constraint, the product of invariants has to be put into correspondence
with a linear expression in the multivector fields. The algebraic
constraint makes this job.}. This means that from the two wavefunctions,
only one represents a new diffeomorphism invariant (that is identified
as the second coefficient of the Alexander-Conway polynomial). This is a
characteristic of the procedure of construction outlined above: in
general the family $\{\psi^{[N,n]}_i\}$ would contain the linear version
of the product of invariants belonging to lower rank families. Other
examples of extended knot families are considered in \cite{Gr3}.

\section{Conclusions}
The basis of a systematic method for obtaining analytic expressions of
diffeomorphism invariants (the extended knots) in term of the
Chern-Simons propagators are settled. As it was shown, any extended knot
would generate an ordinary knot with the only requirement of
substituting the general multivectors by the multitangent fields. The
construction procedure opens a new possibility to explore knots.

This new possibility to look at the world of knots is specially relevant
for quantum gravity. As it is known, the quantum states of gravity in
the loop representation are given by knot invariants (due to the
diffeomorphism invariance of the theory). But the evaluation of the
Hamiltonian constraint onto knot invariants is a very involved task. In
fact, the analysis of the Hamiltonian onto loop dependent wavefunction
has been traditionally based upon geometric (in contrast to analytic)
properties of the loops. The lack of an effective machinery to put
forward the analytical calculations and the limited knowledge of the
analytic properties of knots constitute serious obstacles for the loop
representation of quantum gravity. Extended loops offers a way to
overcome these difficulties. For one hand, the above proposed method
allows to obtain analytic expressions of knot invariants in a generic
way (the Mandelstam identities can also be checked systematically in
this approach). On the other, the explicit analytic evaluation of the
Hamiltonian constraint can be thoroughly accomplished within the
extended loop approach. These features suggest that extended loops could
be an effective resort to advance towards the identification of
the space of states of quantum gravity.

\section*{Acknowledgments}
I want to thank Cayetano Di Bartolo, Rita Gianvittorio, Isbelia Martin
and Alvaro Restuccia for their hospitality during my stay in U. S. B.

\end{document}